# A Speculation into the Origin of Neutral Globules In Planetary Nebulae:

# Could the Helix's Comets Really Be Comets?

Grant Gussie

Department of Physics

University of Tasmania

GPO Box 252C

Hobart, Tasmania

Australia,

7001

email: Grant.Gussie@phys.utas.edu.au

Abstract

A novel explanation for the origin of the cometary globules within NGC 7293 (the "Helix" planetary nebula) is examined; that these globules originate as massive cometary bodies at large astrocentric radii. The mass of such hypothetical cometary bodies would have to be several orders of magnitude larger than any such bodies observed in our solar system in order to supply the observed mass of neutral gas. It is however shown that comets at "outer Oort cloud" like distances are likely to survive past the red giant and asymptotic giant branch evolutionary phases of the central star, allowing them to survive until the formation of the planetary nebula. Some observational tests of this hypothesis are proposed.

The Helix planetary nebula (NGC 7293) is the nearest example of its class, at a distance of 130pc (Daub 1982). The nebula possesses a large and complex multi-shell structure, with the faintest outer filaments visible across a angular diameter of 0.5°, corresponding to a linear diameter of 1.1pc. But even the brightest inner shells of the planetary nebula (hereafter PN), which extend some 12' in diameter, are quite faint, making the Helix a difficult object for spectroscopic study. The central star is also a decidedly faint object, with a luminosity of $600 L_\odot$ (Bohlin et al. 1982), while the central stars of other PNe typically have luminosities $\geq 10000\ L_\odot$. The star's effective temperature is however quite high, at 120000 K (Bohlin et al. 1982). These data are thought to indicate that the nebula is an extremely aged, but massive, PN, and that the central star is in its final evolutionary descent to the white dwarf stage. The dynamical age of the ionised gas is $\gtrsim 10000$ yr, in agreement with this view.

It might therefore be surprising if a molecular component of the original circumstellar material has survived ionisation to the present time. The Helix is however known to possess a molecular envelope, as revealed by $H_2$ (e.g. Storey 1984) and CO observations (e.g. Huggins & Healy 1986). More recent $J = 2 \rightarrow 1$ CO observations of the nebula (Huggins et al. 1992) have indicated that the CO emission is associated with cometary globules within the ionised shell of the nebula, which have long been investigated optically (e.g. Vorontsov-Velyaminov 1968). However, the spatial resolution of the recent CO data was a relatively poor 12", making a firm identification of the CO clumps with the $\approx 1$" optical clumps problematic.

It has however been argued — based on the expected observational properties and lifetimes of such globules — that the cometary globules are indeed cold ($\approx 10$ K) neutral gas in pressure equilibrium with the surrounding $10^4$ K ionised medium (Dyson et al. 1989). Observationally, the cometary globules are *not* seen to be a source of [OIII] emission (which dominates the spectrum of the surrounding ionised shell) but are clearly visible as Hα+[NII] emission (Walsh & Meaburn 1993). This indicates a very much different ionisation state or chemical composition (or both) for the cometary globules and the surrounding ionised shell.

The origin of the cometary globules remains unclear. Traditionally, they have been viewed as condensations within the massive stellar wind of the precursor star, perhaps the remnants of condensations seen as SiO maser spots in evolved stars' envelopes (Dyson et al. 1989). In this paper, a novel origin is however examined: that the cometary globules originate as solid are icy bodies in the process of being ablated by the flowing ionised medium around them. The hypothesis is suggested by the appearance of the globules themselves: they certainly look very much like comets in high resolution optical images, see for example plate 2 of Meaburn et al. (1992).

The mass of the larger cometary globules is however very large; estimated as $\approx 10^{-5}\ M_\odot$ based on pressure-equilibrium requirements (Dyson et al. 1989), on dust absorption measurements (Meaburn et al. 1992), and on CO observations (Huggins et al. 1992). This is obviously much larger than any known solar system comet, being a few times that of the Earth, but the mass distribution function of even the Sun's comets is unknown. Masses of active comets in the inner solar system are very poorly known, but their observed size and expected composition would indicate a typical mass of ~ $10^{-17}$ $M_\odot$. Larger bodies of probably similar composition are however also observed, such as Chiron (~ $10^{-11}$ $M_\odot$) and Pluto (~ $10^{-8}$ $M_\odot$). If active comets, small outer planets, and distant asteroids are part of the same continuous mass distribution of icy bodies, then it would be entirely plausible that a $10^{-5}$ $M_\odot$ body could also exist at the high-mass limit of the distribution, given a large enough population of such objects. The population (or even the very existence) of the Oort Cloud is still undetermined observationally, but empirical estimates place the number of solar system comets at $\approx 10^{13}$ (Weissman 1991). Such a sample could quite easily include at least a few $10^{-5}$ $M_\odot$ objects if one assumes a plausible power law-type mass distribution function for the comets, with a power law dependence of approximately 2. It should however be noted that there is no evidence of such bodies in our solar system indicated by either the gravitational perturbations of the outer planets or the distribution and arrival rate of active comets (presumably of Oort Cloud origin) in the inner solar system. It should however be remembered that the mass of the Helix nebula is several times larger than that of the solar system, making it likely that the outer regions of its stellar system would be more heavily populated with massive bodies than our own.

Assuming that the initial mass function of the Helix's Oort cloud contains comets of the required mass, survival of the comets against outgassing to the present day is quite likely. Outgassing in comets is believed to occur as the result of several different processes. Water ice formed by condensation in the low temperature regions ($\lesssim$ 50K) of the solar nebula should be in an amorphous (non-crystalline) form, and therefore would have trapped a significant amount of volatile gas during condensation (e.g. Bar-Nun et al. 1985). The amorphous ice will then undergo a transformation to a cubic crystalline structure when heated to 125K, and then to the familiar hexagonal crystalline structure when further heated to 160K (Notesco et al. 1991). Both crystal restructurings are accompanied by a release of volatile gases, including CO (Schmitt et al. 1989). Such low-temperature gas release would serve to deplete the comet of CO and more volatile compounds, but leave the mass of the dust and water ice components relatively untouched. Outgassing associated with the active sublimation of water ice is observed only much closer to the Sun, at distances of $\lesssim$ 3AU (e.g. Newburn & Spinrad 1989). To be conservative, an outer sublimation limit of 5AU is adopted here (Weissman 1990). This is *very* conservative, since outgassing of a low-mass comet where gravitational binding is negligible is very different than outgassing in a gravitationally bound $10^{-5}$ $M_\odot$ body, where an escape velocity of ~ 10 km s$^{-1}$ must be obtained by the gas. The Earth does not, after all, have a cometary tail despite being only 1 AU from the Sun. This last statement is however somewhat misleading since the "volatiles" of the Earth (namely its hydrogen and helium) were lost soon after its formation because the thermal velocities of these species exceed the Earth's escape velocity. If the Earth were subjected to the water-dissociating radiation field of a PN nucleus, comet-like mass loss would resume in the form of hydrogen outgassing.

The luminosity of the central star of the Helix is expected to possibly have been has high as 20000 $L_\odot$ in the past (Kippenhahn 1981), causing photo-induced sublimation of only those comets with periastra of ≤ 700 AU; assuming of course that the volatility of extra-solar comets are the same as ours. Because of the pessimistic limits adopted here, the real comets, especially those with larger mass, would likely be far more survivable than this suggests. It is therefore quite clear that the outer Oort cloud comets — and especially our hypothetical high-mass ones — can survive the insolation of a red giant star.

The inner edge of the observed Helix cometary globules is well outside our pessimistic 700 AU limit, lying at ≈ 20000 AU, which is near the canonical distance of the "classical" outer Oort Cloud's inner edge as proposed in the models of many researchers (e.g. Duncan et al. 1987). This indicates that the comets observed in the Helix may represent a population similar in spatial distribution to that proposed for the Sun's Oort cloud. The majority of Oort cloud comets are however now believed to lie well within the classical Oort cloud distance of ≈ 20000 AU, in the so-called inner Oort cloud from 100 AU to 20000 AU (e.g. Weissman 1991) and the Kuiper Belt is now observed (Weissman 1993). There are no cometary globules observed at such small radii in the Helix Nebula, indicating that a mechanism for the destruction of comets with periastra within the range 700 AU < $a$ < 20000 AU is required which leaves comets at $a$ > 20000 AU relatively unaffected. Kwok et al. (1978) proposed an interacting stellar winds model for PN morphology and evolution, where a fast ($\gtrsim$ 2000 km s$^{-1}$) low-mass-loss rate (≈ 10$^{-8}$ $M_\odot$ yr$^{-1}$) stellar wind overtakes and shocks pre-existing wind material from the slow (≈ 20 km s$^{-1}$), high-mass-loss rate (≈ 10$^{-4}$ $M_\odot$ yr$^{-1}$) stellar wind of asymptotic giant branch (AGB) phase of evolution. This sweeps up the now-ionised AGB wind into the dense shells seen in PNe, but also results in a reverse shock that propagates back through the high-velocity wind. The reverse shock creates a high-temperature adiabatic region of ~ 10$^6$ K within the dense ionised shell. This high temperature region is thought to have been observed as X-rays in some PNe (Kreysing et al. 1992) including the Helix (Leahy et al. 1994), where it is located < 20000 AU from the central star. The extreme heating and violent disruption of the high-temperature region would be expected to destroy any comets within them. Only the cooler regions of the ionised shells themselves could be expected to still harbour comets. The inner edge of the cometary globule distribution of Helix is coincident with the inner edge low-ionisation H$\alpha$ emission in images of Walsh & Meaburn (1993) in accordance with this view.

Determination of the exact evaporation lifetime of a comet within the ionised shell is however a very difficult problem because it involves modelling of an ablating, evaporating body within a supersonic flow and an ionising radiation field. Which one of the possible evaporation processes dominates — ablation, evaporation, or photo-dissociation — depends on the thermal conductivity of the cometary material and its structural integrity against ablation, neither of which are known.

The observed properties of the globules (cold, dusty, and molecular) would however require that mechanical ablation by the supersonic wind dominates, since the other mechanisms would imply significant heating and/or dissociation of the released gas. In accordance with this requirement there is evidence (Kouchi et al. 1992) that the thermal conductivity of amorphous vapour-deposited ice is very low, suggesting that the heating of comets is quite inefficient. In any event, heating of solid material in the ionised shells of PNe is known to be ineffectual; for example, the typical IRAS dust temperatures of evolved PNe is $\lesssim 100$ K. The dust temperature of the Helix Nebula (extended IRAS source X2226-210) is only $\approx 50$ K, well below the sublimation temperature of ice. There is also evidence that the surface of comets are quite fragile; Halley's comet was observed to undergo an outburst of activity at a heliocentric distance of 14.3 AU (West et al. 1991), and it has been proposed (Intrilligator & Dryor 1991) that this outburst was due to a collision with particles of a solar flare, which could have imparted a force of only $3 \times 10^{-10}$ N m$^{-2}$ onto the surface. Other explanations for this outburst do however abound (e.g. Hughes 1991). The implied fragility of cometary surfaces would also suggest that ablation of cometary material in the shocked ionised shells of the Helix nebula may occur at a higher rate than thermal conduction of energy into the comet, resulting in cold released gas. This hypothesis does however remain unproven, and only careful modelling of heat diffusion and ablation of icy bodies of the appropriate mass in the environment of planetary nebulae, and of the subsequent acceleration and ionisation of the released gas, may confirm the viability of this hypothesis.

In the meantime, observational tests of this hypothesis are possible. In this model the chemical composition of the cometary globules would be similar to the chemical composition of solar system comets, and be quite unlike the helium and nitrogen-rich (Hawley 1978) ionised shell of the Helix nebula. Determination of the elemental compositions of the globules and ionized gas would therefore test this hypothesis, but the vastly different ionisation states of the cometary globules and the ambient medium makes this very difficult. The chemical composition of neutral gas is not determinable by optical spectroscopy, and therefore infrared and microwave spectroscopic studies of the globules must be made to complement optical observation of the ionised gas. A useful temperature probe would also be found in future $J = 1 \rightarrow 0$ CO observations to complement the $J = 2 \rightarrow 1$ CO observations of Huggins et al. (1992) which will determine the excitation state within the cometary globules.

A proper motion study is another possible test of the comet hypothesis. If gas were being ablated from a cometary body in a distance orbit, ablated gas should move radially away from the cometary head, creating an increasing separation between the head of the stream and individual gas clumps within the stream of 0.02" yr$^{-1}$. High resolution observations of relative clump positions, separated by only a few years, will therefore be valuable in testing the hypothesis.

In addition, this hypothesis predicts that the smaller cometary globules should be more numerous than larger ones to the very limit of the observations resolution – a situation that is observed in Walsh & Meaburn's (1993) photographs. This situation may however be expected to be different at the inner edge of the globules' spatial distribution, where volatile components of the smaller comets would have already evaporated, leaving unobservable dark and dead "cinders" similar to extinct comets in our own inner solar system. An inverse relationship between the size of the smallest globule observable and radius would also provide evidence in support of this hypothesis. Higher resolution optical observations would prove valuable in testing this idea.

Conclusions

Recent high-resolution optical images of the cometary globules in the Helix nebula suggest the intriguing hypothesis that the globules are truly cometary bodies of large mass in the process of ablation. Observational tests of this hypothesis are possible with present technology. However, because of the improbably high mass required of the hypothetical cometary bodies, the conventional explanation for these globules, that they are condensations in the AGB wind, must still be considered the most likely explanation.